\documentclass[]{article}

\usepackage{multirow, amsmath, amsthm, graphicx, amssymb, amsfonts, latexsym, enumerate, amscd, xcolor, float, caption, url}
\usepackage{listings}
\usepackage{geometry}
\RequirePackage[numbers]{natbib}
\usepackage{tcolorbox, subcaption}

\begin{document}
\title{Spatio-temporal insights for wind energy harvesting in South Africa}

\author{Matthew de Bie$^1$, Janet van Niekerk$^{1,2}$, Andri{\"e}tte Bekker$^1$\\
		$^1$Department of Statistics, Faculty of Natural and Agricultural Sciences, University of Pretoria, \\
		South Africa\\
	$^2$Statistics program, CEMSE Division, King Abdullah University of Science and Technology, \\
	Kingdom of Saudi Arabia\\
Janet.vanNiekerk@kaust.edu.sa}

\maketitle

\begin{abstract}
	Understanding complex spatial dependency structures is a crucial consideration when attempting to build a modeling framework for wind speeds. Ideally, wind speed modeling should be very efficient since the wind speed can vary significantly from day to day or even hour to hour. But complex models usually require high computational resources. This paper illustrates how to construct and implement a hierarchical Bayesian model for wind speeds using the Weibull density function based on a continuously-indexed spatial field. For efficient (near real-time) inference the proposed model is implemented in the r package \emph{R-INLA}, based on the integrated nested Laplace approximation (INLA). Specific attention is given to the theoretical and practical considerations of including a spatial component within a Bayesian hierarchical model. The proposed model is then applied and evaluated using a large volume of real data sourced from the coastal regions of South Africa between 2011 and 2021. By projecting the mean and standard deviation of the Mat\'ern field, the results show that the spatial modeling component is effectively capturing variation in wind speeds which cannot be explained by the other model components. The mean of the spatial field varies between $\pm 0.3$ across the domain. These insights are valuable for planning and implementation of green energy resources such as wind farms in South Africa. Furthermore, shortcomings in the spatial sampling domain is evident in the analysis and this is important for future sampling strategies. The proposed model, and the conglomerated dataset, can serve as a foundational framework for future investigations into wind energy in South Africa. 
\end{abstract}

\section{Introduction}\label{intro}
South Africa is facing an electricity crisis which is severely damaging the country's economic output \cite{du2020south}. Ageing infrastructure and a seeming inability to bring new power stations online mean that South Africa has never been more vulnerable to electricity shortages. In a global environment in which the popularity of fossil fuels is waning and positive climate action is viewed as an imperative, South Africa must takes steps towards the implementation of renewable energy initiatives. Harnessing wind energy could provide a solution to South Africa's ongoing energy shortfalls. However, harnessing wind energy is a complex task which requires a nuanced understanding of numerous factors. It is imperative that any analysis aimed at enabling the harnessing of wind energy is grounded in a robust statistical framework.
\par 
Wind energy harvesting in South Africa is becoming a field of increasing interest. Investigations into potentially forecasting wind speeds using machine learning techniques  have been undertaken by \cite{wu2019data} and \cite{dalton2022exogenous}. Additionally, work by \cite{ayodele2019potential} and \cite{ayodele2021optimal} into a connection between wind energy and hydrogen production, as well as the analysis by \cite{mostafaeipour2020statistical} into the viability of using next-generation turbines for generation purposes, indicate that wind energy harvesting is a popular field of research. In light of this popularity, it is unsurprising that efforts have been undertaken by \cite{doorga2022geospatial} to examine some of the spatial considerations associated with harnessing renewable energy resources.
\par 
However, to our knowledge, this paper is the first in South Africa to examine real wind speed data using a hierarchical Bayesian model which includes a spatial modeling component.
\par 
It is recognized by   \cite{garcia1998fitting,zaharim2008suitability,adefarati2019evaluation,ucar2009investigation,seguro2000modern} that the Weibull distribution, the functional form of which is shown in equation \eqref{eqn:weibull}, can serve as a suitable likelihood function for modeling wind speeds.
\begin{eqnarray}
	f\left(y|\alpha,\lambda\right)
	= \frac{\alpha}{\lambda} \left(\frac{y}{\lambda}\right)^{\alpha-1} \exp\left(-\left(\frac{y}{\lambda}\right)\right)^{\alpha}, 
	\label{eqn:weibull}	
\end{eqnarray}
where $\lambda$ is the scale parameter and $\alpha$ is the shape parameter.
We use the Weibull density function given in equation \eqref{eqn:weibull} as the foundation on which to construct a statistical regression model for wind speeds. The core assumptions when working with the Weibull distribution is that $\lambda$ is a function of the covariates, and $\alpha$ is constant \cite{mueller2015constant}. By assuming that $\lambda$ is a function of the predictor variables, the Weibull distribution lends itself towards the construction of a hierarchical regression model.
\par
Our ultimate objective is to include a well defined spatial modeling component within the linear predictor of a hierarchical model in order to gain an understanding of the spatial dependency structures which influence wind speeds. To accomplish this we first describe, in section \ref{chap:data_colect}, how the data relevant to this analysis was collected and cleaned. Thereafter, in section \ref{chap:bayes_hierarch}, we outline the structure of the hierarchical Bayesian model we will fit to the data. This section will also investigate the appropriate priors we will fit to our various modeling components. This will be followed by a discussion, in section \ref{chap:spat}, surrounding how to include a spatial component within the hierarchical modeling structure. Thereafter, section \ref{chap:model} will showcase some of the key results obtained from the model and this will be followed by a short discussion in section \ref{chap:concluion}.

\section{Data collection and conglomeration}\label{chap:data_colect}
\begin{figure}[h]
	\centering
	\includegraphics[scale=0.35]{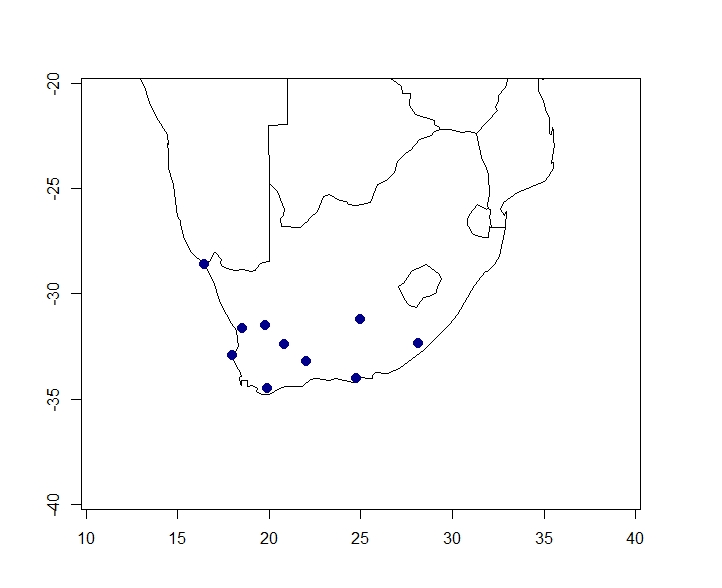}
	\caption{Map of South Africa marked with locations where data was collected.}
	\label{fig:dat_map1}
\end{figure}

To enable our analysis, we constructed a large dataset using raw data sourced from the wind atlas South Africa (WASA) database. Specifically, all raw data was collected from measuring sites in the coastal regions of South Africa between 2011 and 2021. The details of the locations where the raw data was collected are shown in \ref{fig:dat_map1} and \ref{table:site_coord1}. Between 2011 and 2021, the wind speed was recorded every five minutes, at five different altitude levels, and at ten different locations. This collection process resulted in a large amount of disparate data which needed to be organised into a single coherent dataset which is capable of supporting our analysis.
\par 
After the conglomeration process was completed, the final dataset contained over $25 000 000$ observations, hereafter we shall refer to this dataset as the prime dataset. Each observation in the prime dataset was identified by the variables shown in \ref{table:newvar1}.
\begin{table}[h]
		\centering
	\begin{tabular}{||l|l|l|l||}	\hline
		Station\_ID & Site Name & Latitude & Longitude 	 \\ \hline\hline
		WM01 & Alexander Bay & -28.583331 & 16.4833 \\
		WM02 & Calvinia & -31.4707 & 19.7760 \\
		WM03 & Vredendal & -31.6391 & 18.5285 \\
		WM04 & Vredenburg & -32.9000& 17.9833 \\
		WM05 & Napier & -34.4667 & 19.9000 \\
		WM06 & Sutherland & -32.3743 & 20.8064 \\
		WM07 & Prince Albert & -33.2167 & 22.0333 \\
		WM08 & Humansdorp & -34.0027 & 24.7440 \\
		WM09 & Noupoort & -31.1874 & 24.9499 \\
		WM10 & Butterworth & -32.3308 & 28.1498 \\
		\hline
	\end{tabular}
	\caption{Coordinates of sites at which data was recorded.}\label{table:site_coord1}
\end{table}
It must be noted that two sites, WM04 and WM10 suffered malfunctions in the mid-2010s which meant that they were unable to record data.
\begin{table}[h]
		\centering
	\begin{tabular}{|l|l|}	
		\hline
		Variables &	wind\_speed, date\_time, station\_ID, latitude, longitude, month, year,\\
		&  altitude, wind\_direct AVG, cos\_direct, sin\_direct, f\_month, c\_month \\
		\hline
	\end{tabular}
	\caption{Variables in conglomerated dataset.}\label{table:newvar1}
\end{table}

\par
Having constructed a usable dataset, we now need to investigate the hierarchical Bayesian framework we will utilise to fit a model to this data.

\section{A hierarchical Bayesian modeling framework}\label{chap:bayes_hierarch}
In order to fully understand a hierarchical Bayesian regression model, it is useful to break the term down into its component words. 
\par  
The term hierarchy indicates that we are constructing a model which has multiple levels. Furthermore the word Bayesian denotes that we are working with priors to estimate posterior density functions. Finally, the word regression indicates that we are building a model which aims to explain the relationship between a matrix of covariates $\mathbf{X}$ and a response variable, $\mathbf{y}$.
\par  
Therefore, we understand that we are working with a multi-level model where we attach priors to the various components at each level of the model. Using these priors, we will obtain a posterior density function for each parameter. With this in mind, we can now specify the manner in which each level of this model will function.
\par 
With reference to \ref{table:newvar1}, we designate 'wind\_speed' as the response variable, $\mathbf{y}$. Additionally, we assume that there exists some form of relationship between the response and the other variables listed in table \ref{table:newvar1}. We designate these variables as our matrix of covariates, which we label as $\mathbf{X}$. Alongside standard fixed effect, our model will also make use of random effects in the latent field. Notationally, $\mathbf{Z}$ acts as a design matrix which allows a single observation to depend on multiple random effects and $\mathbf{u}$ serves as the vector which contains the random effects.
\par 
The final aspect to be aware of before we outline the modeling framework is that the unknown prior density function for hyperparameters, like the shape parameter $\alpha$, which have an unknown functional form will be written as $\alpha \sim H(...)$.

\subsection{General form of a hierarchical Bayesian model}\label{chap:bayes_hierarch_general}
The general form of a Bayesian hierarchical modeling framework, constructed with reference to the Weibull density function , is shown below.
\begin{enumerate}
	\item Likelihood function: $\mathbf{y} \sim \text{Weibull}(\lambda, \alpha)$
	\item Link function to the mean: $\lambda= \exp(\eta)$
	\item Linear predictor $\eta = \mathbf{X} \mathbf{\beta} +\mathbf{Z}\mathbf{u}$
	\item Hyperparameter: $\alpha \sim H(...)$.
\end{enumerate} 
It is within $\eta$ that we will specify the regression relationship between the covariates $\mathbf{X}$ and the response variable, $\mathbf{y}$. This structural form also allow us to define random effects within the linear predictor. Next, we need to identify a suitable prior for $\alpha$.

\subsection{Identifying a suitable prior for $\alpha$}\label{chap:hierarch_prior}
It was recognised by \cite{van2021principled} that popular existing priors for the shape parameter of the Weibull distribution, such as the improper uniform prior and the gamma prior, are often unsuitable when performing complex modeling. With this in mind, the work of \cite{simpson2017penalising} was used as a basis by \cite{van2021principled} to derive a penalised complexity (PC) prior for the shape parameter of the Weibull distribution.
\par 
Using the Kullback-Liebler divergence, \cite{van2021principled} gives a full definition of the PC prior for the Weibull shape parameter as formulated in equation \eqref{eqn:PC1}.
\begin{eqnarray}
	\pi(\alpha)&=& \theta \exp[-\theta \sqrt{2KLD(\alpha)}] \left|\frac{\partial\sqrt{2KLD(\alpha)}}{\partial \alpha}\right| \notag \\
	&=& \theta \exp[-\theta\sqrt{2KLD(\alpha)}] \times \frac{1}{2}(2KLD(\alpha))^{-\frac{1}{2}} \notag \\
	&\times& \left |\frac{2}{\alpha^{2}}(\gamma-\alpha(1+\gamma))
	+\Gamma(\alpha^{-1}+\alpha\log(\alpha)) \right.
	\notag \\ 
	&+& \left.\frac{2}{\alpha} \left(-(1+\gamma)-\frac{\Gamma(\alpha^{-1})\psi (\alpha^{-1})}{\alpha^{2}}+\log(\alpha)+1\right)\right|, \notag \\ \label{eqn:PC1}
\end{eqnarray}
where $\psi(z)$ is the digamma function as defined by \cite{abramowitz1988handbook}. In a practical sense, the PC prior allows us to attach generally applicable priori information to $\alpha$. In other words, the PC prior is a useful default prior for $\alpha$ in the absence of other strong prior information.

\subsection{Specifying a model within the hierarchical framework}
We can now specify our mixed effect regression model using the general structural form defined in \ref{chap:bayes_hierarch_general}.
\begin{eqnarray}
	Y &\sim& \text{Weibull}(\lambda, \alpha) \notag \\
	\lambda &=& \exp(\eta) \notag \\
	\eta &=& \beta_{0} + \beta_{\mathtt{cos}} \mathbf{X_{\text{cos\_direct}}} + \beta_{\mathtt{sin}}\mathbf{X_{\text{sin\_direct}}}+ g(\mathbf{X_{\text{Altitude}}}) \notag \\
	&+& v(\mathbf{X_{\text{f\_month}}}) + q(\mathbf{X_{\text{c\_month}}}) + \mathbf{u_{s}} \notag \\
	\alpha &\sim& \text{PC prior} , \label{eqn:model_h1}
\end{eqnarray}
where, $g(\mathbf{X_{\text{Altitude}}})$, denotes that we use a spline component, of order random walk 2, in order to capture the variation in wind speeds which arises from the different altitudes at which the wind speed was measured. Additionally, $v(\mathbf{X_{\text{f\_month}}})$ and $q(\mathbf{X_{\text{c\_month}}})$ serve as the temporal effects within the model. Both $v(\mathbf{X_{\text{f\_month}}})$ and $q(\mathbf{X_{\text{c\_month}}})$ are auto-regressive (AR)1 random effects. Finally, two fixed effects, namely $\beta_{\mathtt{cos}} \mathbf{X_{\text{cos\_direct}}}$ and $\beta_{\mathtt{sin}}\mathbf{X_{\text{sin\_direct}}}$ are used to capture the impact of the direction on the wind speed.
\par 
In addition, we have defined two temporal effects, namely $v(\mathbf{X_{\text{f\_month}}})$ and $q(\mathbf{X_{\text{c\_month}}}) $. We refer to the list of variables provided in \ref{table:newvar1} to provide reasoning as to why we included two temporal effects in our model. Within \ref{table:newvar1} there are two variables which are temporal in nature 'f\_month' and 'c\_month'. The first, 'f\_month' serves as a repeating temporal component which is intended to capture cyclical temporal dependency which occurs year-on-year. In other words, every year the wind speed in February will depend on the wind speed in January. This latent variation could be caused by repeating seasonal patterns. Therefore, we use component $v(\mathbf{X_{\text{f\_month}}})$ in order to capture a dependency structure which will repeat each year. 
\par
Conversely, the wind speed in January of a given year, also has a relationship with the wind speed in December of the previous year. To capture temporal relationships across each of the ten years, we define component $q(\mathbf{X_{\text{c\_month}}})$.
\par  
The last component specified in the latent field is defined simply as $\mathbf{u_{s}}$ which serves as the spatial component of the model.

\section{Using a Matérn field to capture spatial variation in wind speeds}\label{chap:spat}
Though it is not visible in equation \eqref{eqn:model_h1}, within $\mathbf{u_{s}}$ are a number of estimable parameters which relate to the Matérn correlation structure which governs the manner in which we conceptualise dependency between two points. Using the work of \cite{krainski2018advanced} we can define $\mathbf{u_{s_{i}}}$ where $i=1,2,...,n$ as a realisation of the random spatial effect at $n$ locations within the domain. Assuming that $\mathbf{u_{s}}$ has a multivariate Gaussian distribution and that $\mathbf{u_{s}}$ is continuous across space, we can then conceptualise $\mathbf{u_{s}}$ as a continuously-indexed Gaussian field (GF).
\par 
It was suggested by \cite{rue2002fitting} that it is theoretically possible to approximate a GF with a Gauss-Markov random field (GMRF). From this hypothesis, \cite{lindgren2011explicit} defined a practical method whereby a stochastic partial differential equation (SPDE), the solution to which is a GF with Matérn correlation, could be used to create a GMRF approximation of a GF. Before we move forward, it is useful to briefly review the Matérn covariance function.
\par  
The Matérn covariance function can be used to define the covariance between two points separated by some form of spatial distance. According to \cite{krainski2018advanced} a standard Matérn covariance function can be written as 
\begin{eqnarray}
	Cor_{M}(u_{s_{i}},u_{s_{j}}) = \frac{2^{1-\nu}}{\gamma(\nu)} (\kappa||s_{i}-s_{j}||)^{\nu}K_{\nu}(\kappa||s_{i}-s_{j}||),
	\label{eqn:mcov}
\end{eqnarray}
where $||...||$ denotes the Euclidean distance between $s_{i}$ and $s_{j}$, $\kappa$ is the scale parameter, $v$ is the smoothness parameter, and $K_{v}$ is the modified Bessel function. The logic which underpins equation \eqref{eqn:mcov} states that the degree of dependence between two points $u_{s_{i}},u_{s_{j}}$ is determined by the Euclidean distance between the points and a number of estimable parameters.
\par 
According to the work of \cite{krainski2016r} there are many situations in which we assume that we are working with an underlying GF but cannot directly observe the GF. Instead we observe data with a measurement error. This can be expressed this mathematically as given in equation \eqref{eqn:nugget}.
\begin{eqnarray}
	k_{i} &=& x_{i} + e_{i} \notag \\
	\text{where } e_{i} &\sim& N(0, \sigma^{2}_{e}) , \label{eqn:nugget}
\end{eqnarray}
where $i$ for $i=1,2,...,n$ is the location at which we observe data point $k_{i}$, and $\sigma^{2}_{e}$ measures the noise of the process, otherwise known as the 'nugget' effect.
\par
An aspect of the Matérn field which is not immediately visible from equation \eqref{eqn:mcov} is the range of the correlation effect, often referred to as the nominal range. The nominal range is defined as  the distance at which the correlation, given in equation \eqref{eqn:mcov}, is equal to 0.05 \cite{lindgren2015bayesian}. The true value of the nominal range can be determined by using the formula show in equation \eqref{eqn:nom_ran} \cite{krainski2016r}.
\begin{eqnarray}
	\text{nominal range}&=&\frac{\sqrt{8 \nu}}{\kappa}, \label{eqn:nom_ran}
\end{eqnarray}
where $\nu$ is drawn from equation \eqref{eqn:mcov}. With all of the elements of the Matérn field now defined, we can move on to examining how \cite{lindgren2011explicit} created a GMRF approximation of this specific GF.
\par  
The study by \cite{lindgren2011explicit} found two crucial results. The first of these critical results is that a GF $\mathbf{u_{s}}$ with Matérn covariance is a solution to the linear fractional SPDE shown in equation \ref{heat}.
\begin{eqnarray}
	(k^{2}-\delta)^{\Upsilon /2} \mathbf{u_{s}} = \mathbf{W(s)}, \label{heat}
\end{eqnarray}
where $\delta$ denotes what is known as the  Laplacian operator, $\mathbf{W(s)}$ is spatial random Gaussian white noise, and $\Upsilon$ governs the smoothness parameter of the process.
\par  
The second result found by \cite{lindgren2011explicit} was achieved using the finite element method (FEM) and provides a solution for points distributed on an irregular grid. When we typically assess spatial dependence, most points are not located at regular grid intervals but are rather distributed irregularly across the domain. It was found by \cite{lindgren2011explicit} that the domain can be approximated using non-intersecting triangles. In statistical terms, this irregular triangular grid is known as a constrained refined Delaunay triangulation (CRDT). The result of approximating the domain using the CRDT is a mechanism for spatial analysis called a mesh.

\section{Model implementation and results}\label{chap:model}

\subsection{Data preparation}\label{sect:data_prep}
\begin{figure}[h]
	\begin{center}
		\includegraphics[scale=0.35]{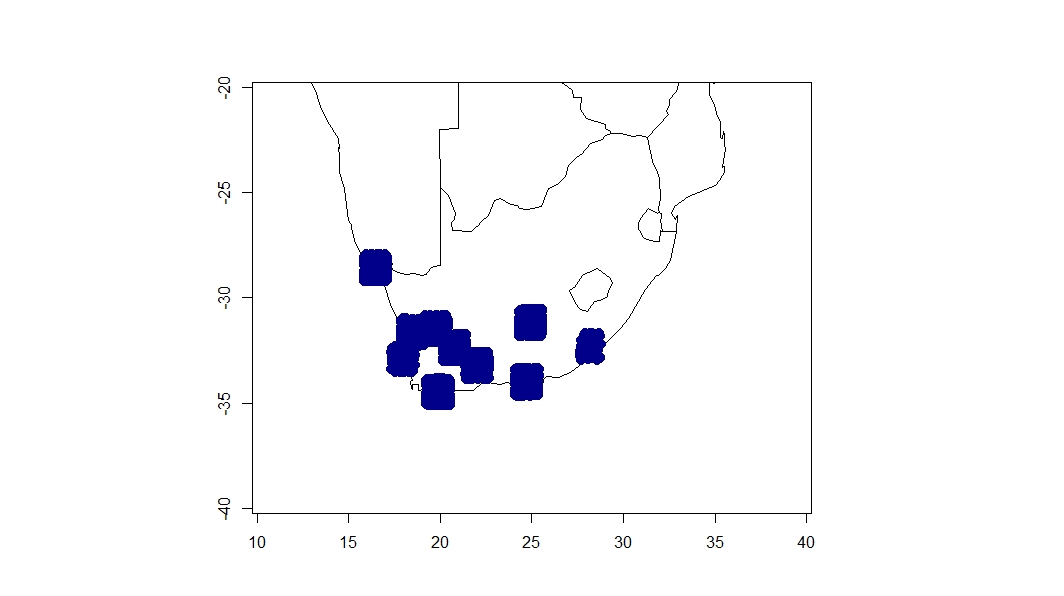}
		\caption{Map of South Africa. Observations are jittered around each measuring site.}
		\label{fig:sa_map_jit}
	\end{center}
\end{figure}

Considering that the data was effectively only recorded at ten locations, we manipulated the data in order to induce a degree of spatial diversity into the dataset. This was achieved by jittering the coordinates in such a manner that every observation technical occupies a unique location. \ref{fig:sa_map_jit} is used to illustrate how this jittering was performed using a sample of $5000$ observations.

\subsection{Mesh selection}
A crucial consideration when implementing a Mat\'ern field, is the selection of an appropriate mesh. However, as noted by \cite{righetto2020choice}, at present there is no firm set of criteria which can be used to precisely select an ideal mesh. In a practical sense, it is possible to identify a suitable as the mesh at which estimated posterior density functions associated with the Mat\'ern field converge. It may be tempting to bypass this selection process by simply using a very dense mesh in order to perform modeling. Considering that we are working with a GMRF approximation of a GF the decision to use a very dense mesh may, at first instance, seem logical. However, from a practical standpoint, the accuracy of estimated model parameters may be seriously impaired due to the computational instability which can arise when using a very fine mesh.
\par
\begin{figure}[h!]
	\centering
	\begin{subfigure}{0.22\textwidth}
		\includegraphics[width=4cm]{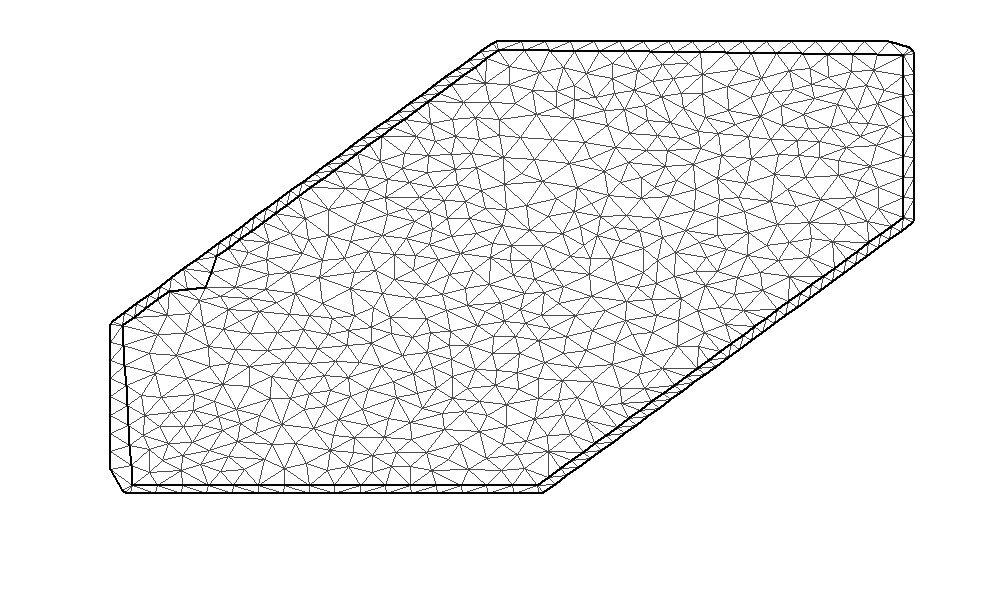}
		\caption{Mesh-A}
	\end{subfigure}
	\begin{subfigure}{0.22\textwidth}
	\includegraphics[width=4cm]{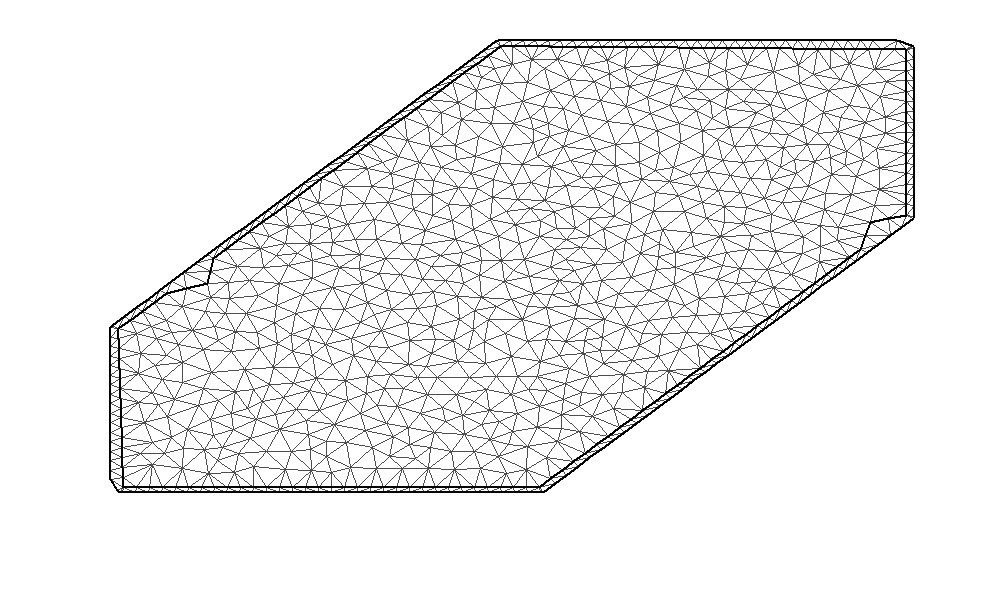}
			\caption{Mesh-B}
\end{subfigure}
	\begin{subfigure}{0.22\textwidth}
	\includegraphics[width=4cm]{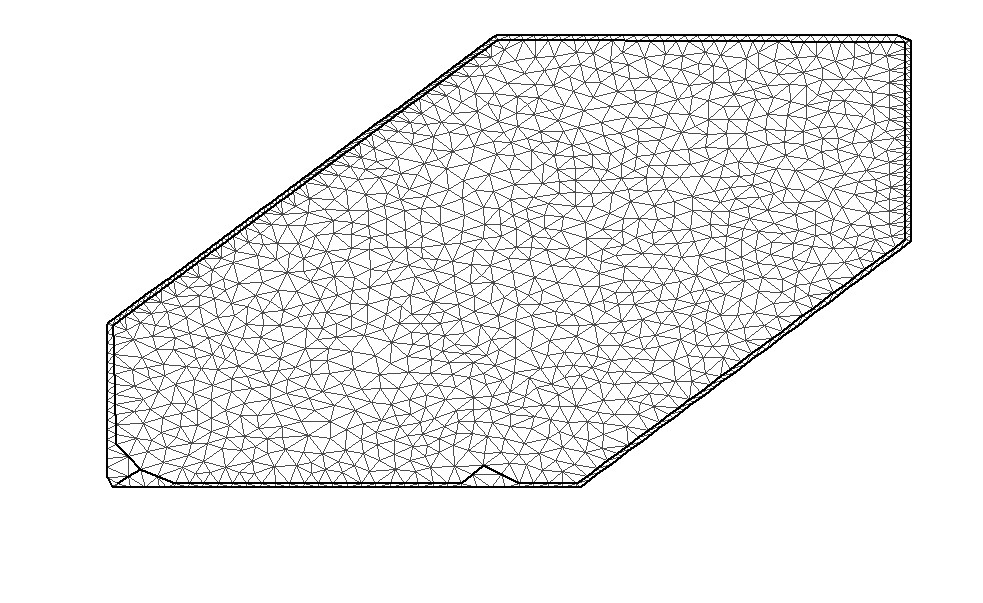}
				\caption{Mesh-C}
\end{subfigure}
	\begin{subfigure}{0.22\textwidth}
	\includegraphics[width=4cm]{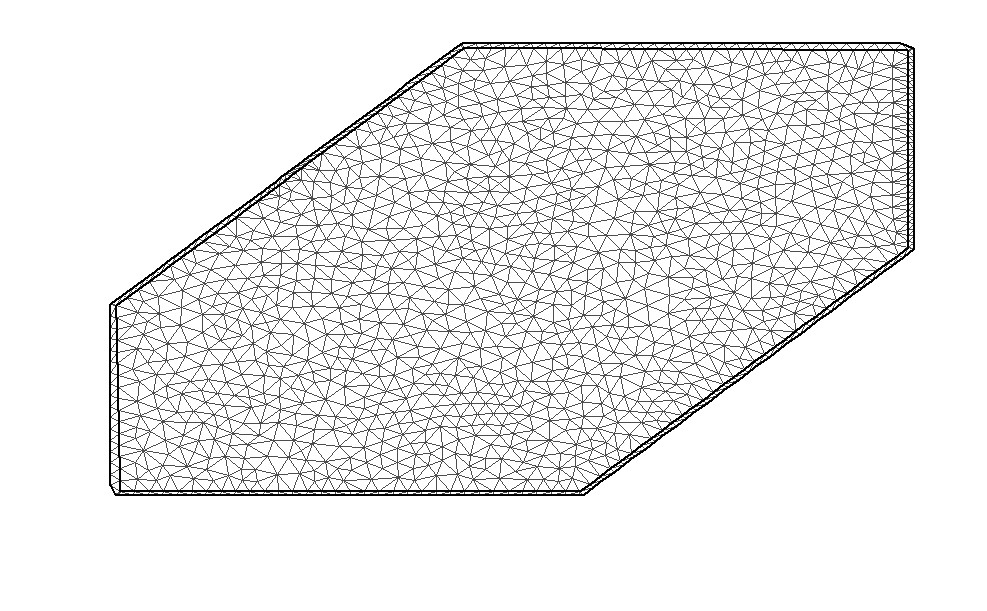}
				\caption{Mesh-D}
\end{subfigure}
	\begin{subfigure}{0.22\textwidth}
	\includegraphics[width=4cm]{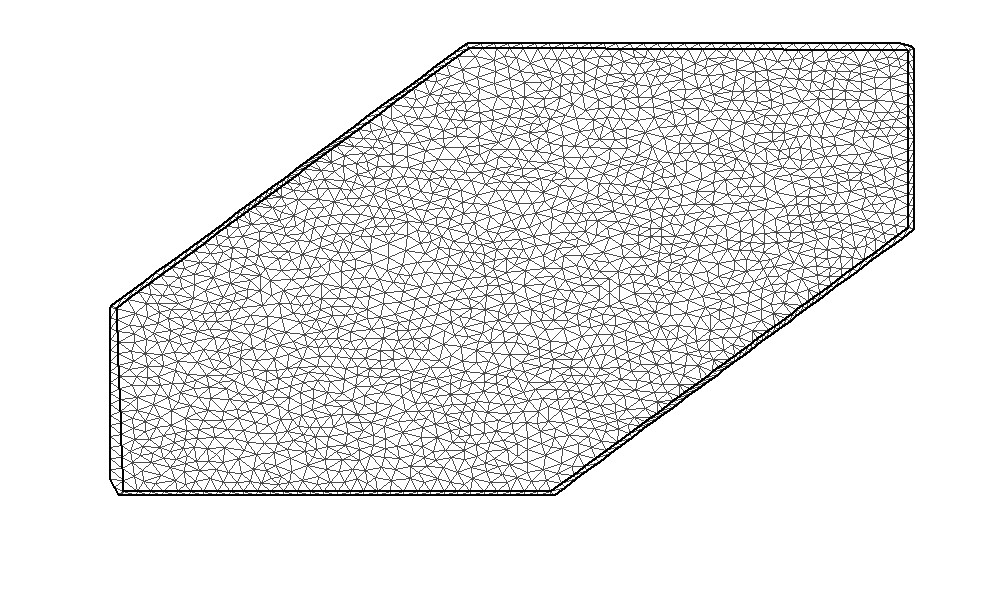}
				\caption{Mesh-E}
\end{subfigure}
	\begin{subfigure}{0.22\textwidth}
	\includegraphics[width=4cm]{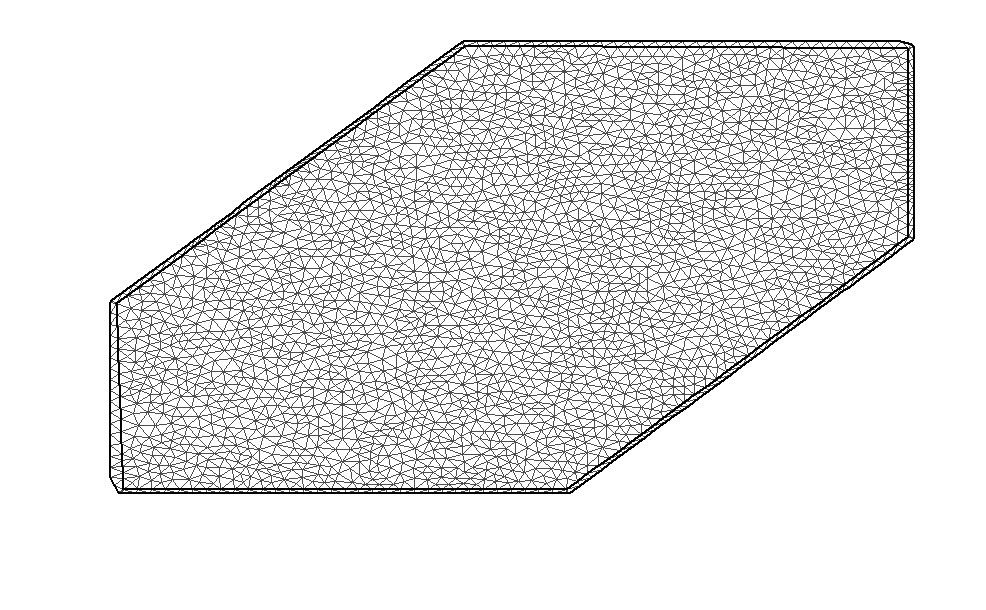}
				\caption{Mesh-F}
\end{subfigure}
	\begin{subfigure}{0.22\textwidth}
	\includegraphics[width=4cm]{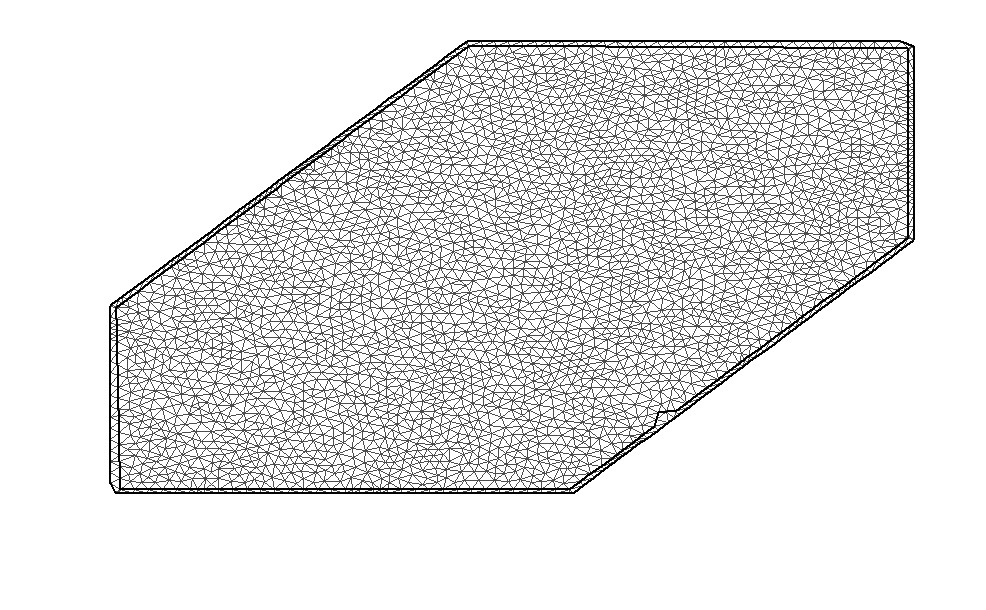}
				\caption{Mesh-G}
\end{subfigure}
	\begin{subfigure}{0.22\textwidth}
	\includegraphics[width=4cm]{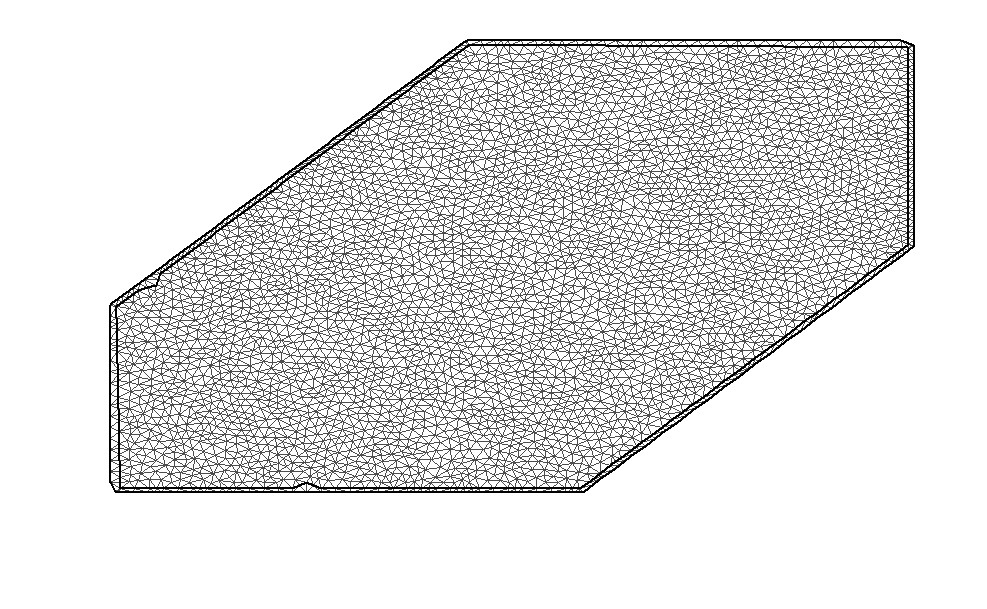}
				\caption{Mesh-H}
\end{subfigure}

	\label{fig:res_mesh1}
\end{figure}

 
The process of mesh selection involves performing a number of redundant calculations. This is because we  need to fit our model using a given mesh and then analyse the results obtained from said mesh in order to determine which mesh is most suitable. Therefore, for purposes of computational efficiency, the process of mesh selection was performed using a sample of $5000$ observations drawn from the prime dataset. The meshes which were generated for purposes of evaluation are shown in \ref{fig:res_mesh1}.

\begin{table}[h]
		\centering
	\begin{tabular}{||l|l|l|l|l|l|l|l||}
		\hline
		Mesh& $me_{1}$ &  $me_{2}$ & $of_{1}$ &$of_{2}$ & $cu_{1}$  & Vertices & CPU time  \\ 
		\hline \hline
		Mesh-A &$1$ &  $1$ & $0.30$ &$0.30$ & $0.95$  & 762 & 15.04 \\
		
		Mesh-B &$0.90$ &  $0.90$ & $0.20$ &$0.20$ & $0.90$ & 981 & 15.67  \\
		
		Mesh-C &$0.75$ &  $0.75$ & $0.15$ &$0.15$ & $0.75$ & 1358 & 27.24 \\
		
		Mesh-D &$0.725$ &  $0.725$ & $0.15$ &$0.15$ & $0.725$ & 1420 & 45.63\\
		
		Mesh-E &$0.55$ &  $0.55$ & $0.15$ &$0.15$ & $0.55$ & 2194 & 49.98\\
		
		Mesh-F &$0.50$ &  $0.50$ & $0.15$ &$0.15$ & $0.50$ & 2561 & 40.76  \\
		
		Mesh-G &$0.45$ &  $0.45$ & $0.15$ &$0.15$ & $0.45$ & 3136 & 50.93  \\
		
		Mesh-H &$0.40$ &  $0.40$ & $0.15$ &$0.15$ & $0.40$ & 4009 & 469.22 \\
		\hline
	\end{tabular}
	\caption{From left to right: maximum.edge.1 ($me_{1}$), maximum.edge.2($me_{2}$), offset.1 1($of_{1}$), offset.2($of_{2}$), number of vertices in the mesh, time in seconds required to use the mesh within the INLA framework}\label{table:meshtest1}
\end{table}

\begin{table}[h]
		\centering
	\caption{From left to right: $\sigma^{2}_{e}$, $\sigma^{2}_{x}$, $\kappa$, nominal range, $\tau$}\label{table:meshtest2}
	\begin{tabular}{||l|l|l|l|l|l||}
		\hline
		Mesh&  $\sigma^{2}_{e}$ & $\sigma^{2}_{x}$ & $\kappa$ & nr & $\tau$ \\ 
		\hline \hline
		Mesh-A & $0.4584$ & $0.0130$ & $0.5478$ & $0.4129$ & $0.1572$\\
		
		Mesh-B & $0.4611$ & $0.0085$ & $0.3677$ & $0.5720$ & $0.2848$\\
		
		Mesh-C & $0.4579$ & $0.0095$ & $0.3466$ & $0.4744$ & $0.2835$\\
		
		Mesh-D & $0.4585$ & $0.0079$ & $0.2703$ & $0.4898$ & $0.2910$\\
		
		Mesh-E & $0.4588$ & $0.0076$ & $0.3426$ & $0.4972$ & $0.2540$\\
		
		Mesh-F & $0.4620$ & $0.0071$ & $0.2707$ & $0.5005$ & $0.3455$\\
		
		Mesh-G & $0.4572$ & $0.0086$ & $0.3003$ & $0.5020$ & $0.2899$\\
		
		Mesh-H & $0.4621$ & $0.0067$ & $0.1734$ & $0.4096$ & $0.3293$\\
		\hline
	\end{tabular}
\end{table}
\par 
In \ref{table:meshtest1} we provide the parameter values which were used along with the \textit{inla.mesh.2d()} function in order to generate each of the meshes shown in \ref{fig:res_mesh1}. Thereafter the estimated point estimates for the Mat\'ern parameters are shown in \ref{table:meshtest2}.
\par
Mesh-E appears to be the mesh at which convergence occurs for most if not all of the parameters in the Mat\'ern field. Additionally, Mesh-E required under fifty seconds in order to be processed within our modeling framework and contains over two thousand vertices. Conversely, the second most suitable mesh, Mesh-D, contains less than one thousand-five hundred vertices and could be too sparse for purposes of parameter inference.
\par 
For our purposes, we would identify Mesh-E as the optimal mesh to use in further stages of this reproducible example. However, if the specific properties of Mesh-E are not in line with those desired by the user then Mesh-D could be used as an alternative suitable spatial mesh.

\subsection{Key model results}\label{sect:results}
Using the integrated nested Laplace approximation  (INLA) we fit the model defined in equation \eqref{eqn:model_h1} to a sample of $1 000 000$ observations drawn from the prime dataset. Utilising an approximate Bayesian method like INLA to fit the model is far more computationally efficient than the computational burden which would be incurred when fitting the model using a Markov-chain Monte Carlo (MCMC) method.
\par
\begin{figure}[h!]
	\centering
	\begin{subfigure}{0.45\textwidth}
		\includegraphics[width=7cm]{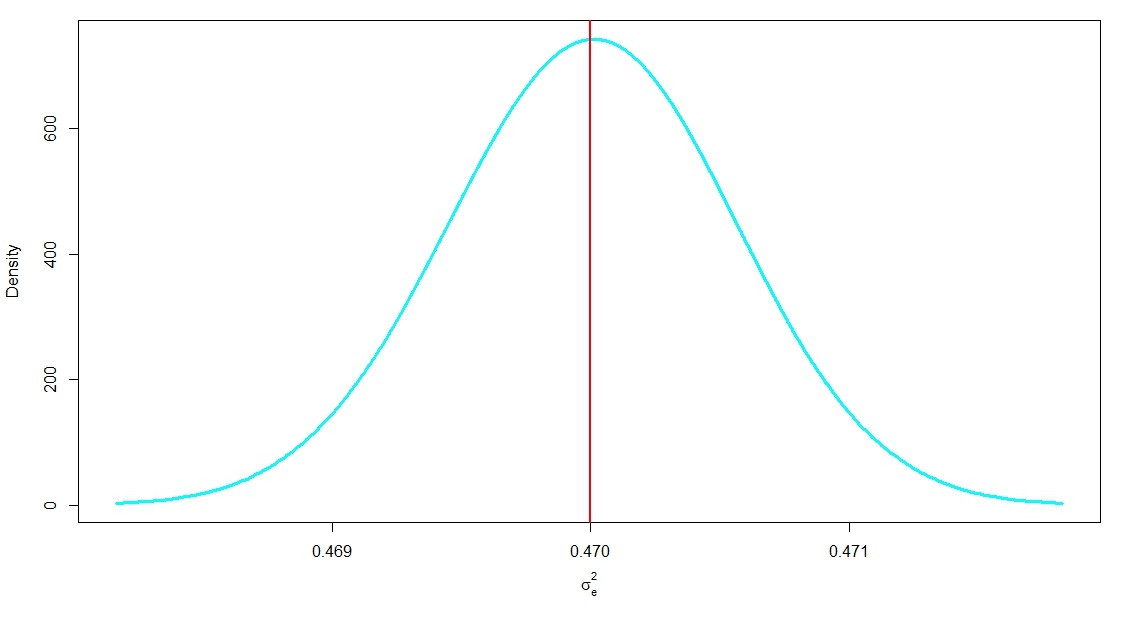}
			\caption{Estimated posterior for $\sigma^{2}_{e}$}
	\end{subfigure}
	\begin{subfigure}{0.45\textwidth}
	\includegraphics[width=7cm]{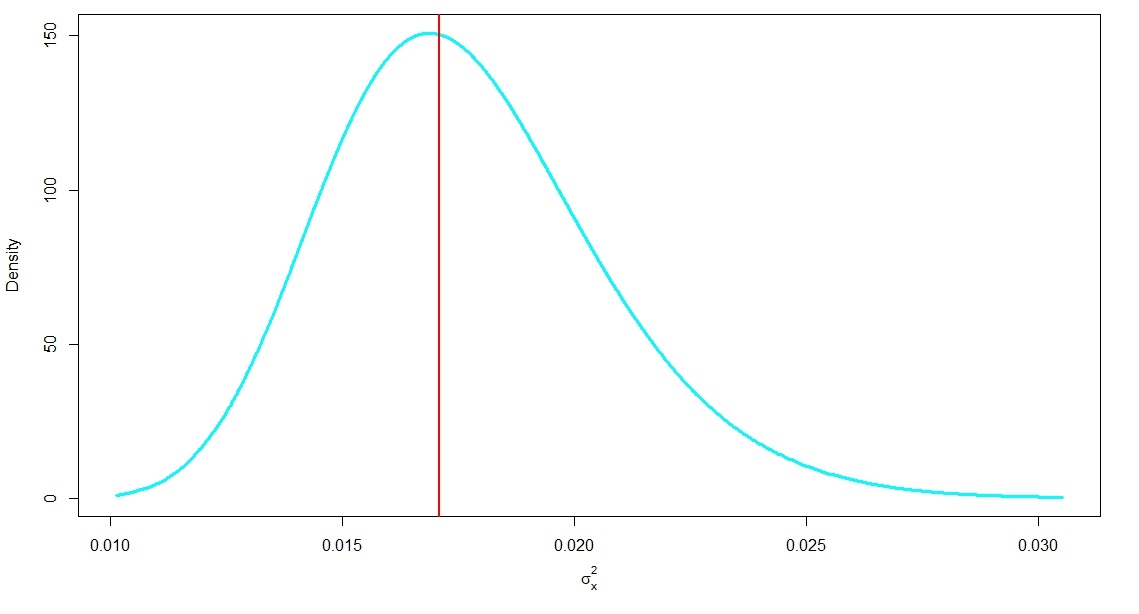}
	\caption{Estimated posterior for $\sigma^{2}_{e}$}
\end{subfigure}
	\begin{subfigure}{0.45\textwidth}
	\includegraphics[width=7cm]{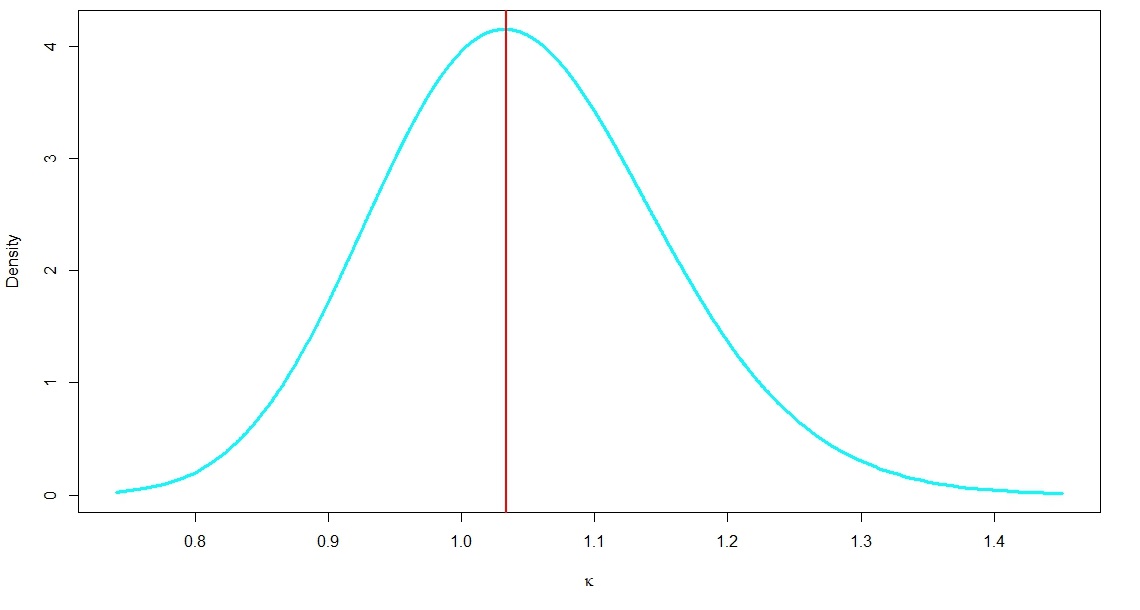}
	\caption{Estimated posterior for $\kappa$}
\end{subfigure}
	\begin{subfigure}{0.45\textwidth}
	\includegraphics[width=7cm]{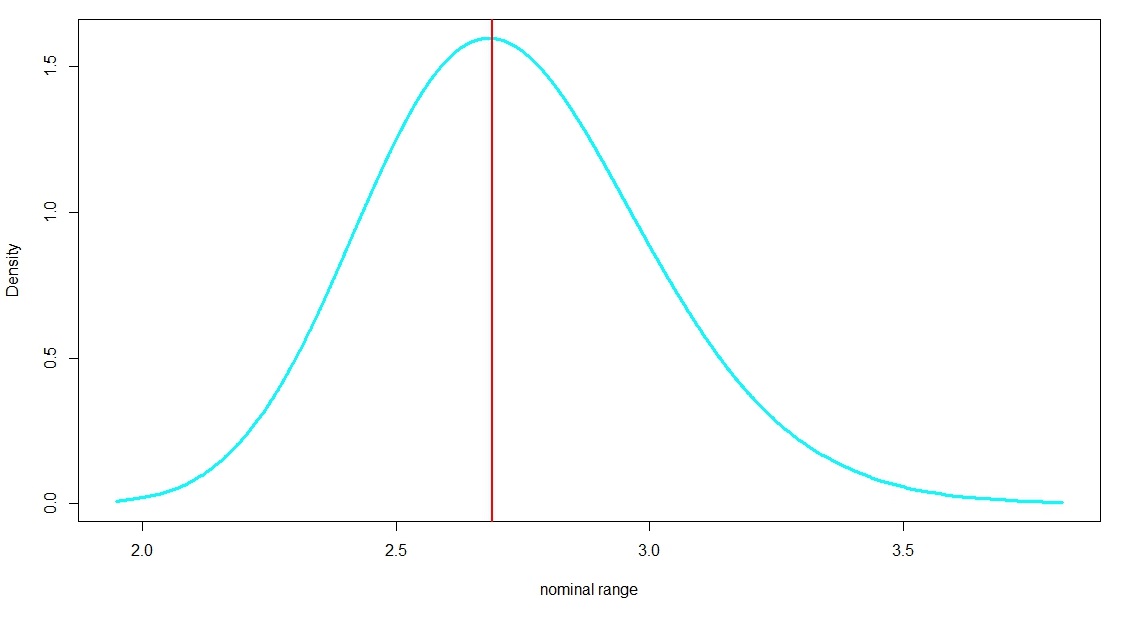}
	\caption{Estimated posterior for nominal range}
\end{subfigure}
	\begin{subfigure}{0.45\textwidth}
	\includegraphics[width=7cm]{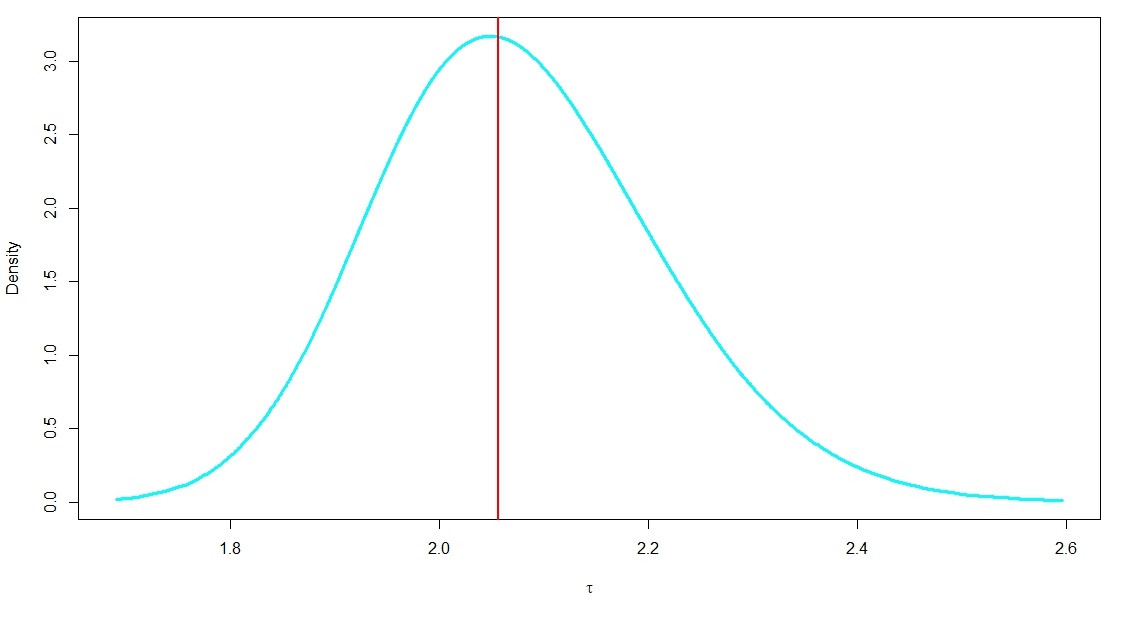}
	\caption{Estimated posterior for $\tau$}
\end{subfigure}
	\label{fig:sm1_pres1}
\end{figure}
\begin{figure}[h!]
	\centering
	\includegraphics[width=12cm]{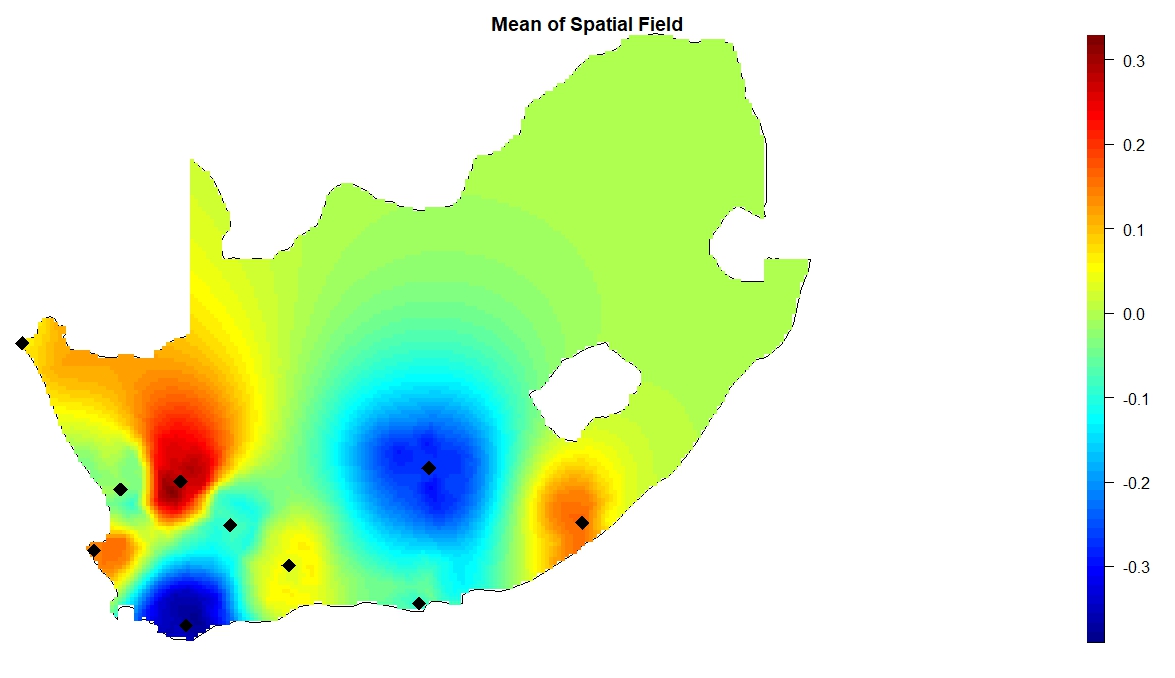}
	\caption{Visualising the mean of the spatial field.}
	\label{fig:psample_spat_mean}
\end{figure}
\begin{figure}[h!]
	\centering
	\includegraphics[width=12cm]{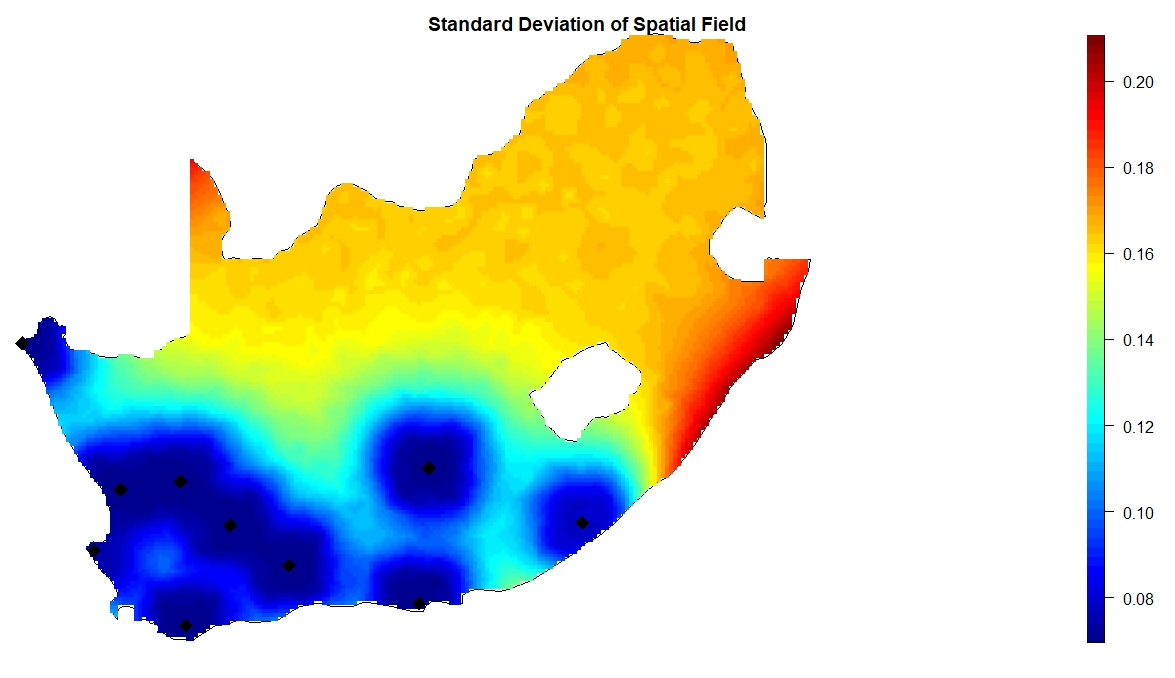}
	\caption{Visualising the standard deviation of the spatial field.}
	\label{fig:psample_spat_sd}
\end{figure}

In \ref{fig:sm1_pres1} and \ref{table:P_res1} we report the posterior density functions and the point estimates for the parameters which govern the spatial field. However, when viewed from an abstract perspective, the estimates in \ref{table:P_res1} are difficult to interpret. With this in mind, we project the mean and standard deviation of the spatial field onto a map of South Africa and visualise these results using \ref{fig:psample_spat_mean} and \ref{fig:psample_spat_sd} respectively.
\par
A comparison of the results in \ref{table:meshtest2} and \ref{table:P_res1} will reveal substantial differences in the point estimates of the parameters of the Matérn field. This is due to the fact that the results in \ref{table:P_res1} were obtained using a greater volume of data, $1000000$ obervations as oppose to $5000$ observations. Additionally, it must be noted that, while the sample of $5000$ observations is useful for preliminary and exploratory analysis, the reduced volume incurred when working with only $5000$ observations can lead to inconsistent results. For these reasons, we chose to report the results obtained when fitting the model to a sample which contained a larger volume of data.

\begin{table}[h]
	\centering
	\begin{tabular}{|l|l|}
		\hline
		Parameter in the Matérn field & Point Estimate \\ 
		\hline
		$\sigma^{2}_{e}$ &  $0.4700$\\
		$\sigma^{2}_{x}$ &  $0.0171$ \\
		$\kappa$ & $1.0341$ \\
		Nominal Range & $2.6891$ \\
		$\tau$ & $2.0562$  \\
		\hline
	\end{tabular}
	\caption{Estimates obtained for the parameters of the Matérn field.}\label{table:P_res1}
\end{table}

The projection of the mean of the spatial field shown in \ref{fig:psample_spat_mean} allows us to visualise how wind speed varied across South Africa once all other model covariates were taken into account. The 'red' regions shown in \ref{fig:psample_spat_mean} indicate an area in which the measured wind speeds were greater than the other model effects. Using the same logic we can therefore understand that a 'blue' region denotes an area in which the measured wind speeds were less than what we would believe the wind speed to be based only on the other model effects.
\par
An examination of \ref{fig:psample_spat_mean} reveals that the mean of the Mat\'ern field is zero across large parts of the domain. This occurs because of the lack of spatial diversity in the underlying data. Strictly speaking, there are only ten locations where recordings of wind speeds were made and the spatial modeling component cannot measure the relationship between wind speeds and location in areas where no data exists. 
\par 
\ref{fig:psample_spat_sd} provides further insight into the manner in which our spatial modeling component performed by projecting the standard deviation of the spatial field over the map of South Africa. Visually, we note from \ref{fig:psample_spat_sd} that the standard deviation is low in the South-West of the country but increases as we move East and North. This occurs because the majority of the measuring sites are located in the South-West of South Africa. Practically, this means that we are more certain of the predicted mean of our spatial field in the South-West regions and become less certain of the predicted mean as we move away from these regions.
\par  
A closer assessment of \ref{fig:psample_spat_mean} reveals that the spatial variation in wind speeds is fairly minor and oscillates between -0.3;0.3. What this demonstrates is that most of the variability in wind speeds is effectively captured by the other covariates in our model. However, the fact that there is any variability at all indicates that the spatial component is effectively capturing latent variation which is not accounted for by the other components in the model. 

\section{Discussion and conclusion}\label{chap:concluion}
In this paper, we constructed a hierarchical Bayesian model and fitted this model to wind speed data sourced from the coastal regions of South Africa. The spatial variation in wind speeds was captured using a Mat\'ern field. Using the INLA method, the results of the model, fitted to a sample of $1000000$ observations, were projected over the domain of South Africa.
\par 
A simple logical assumption proposes that geographical conditions affect the speed of the wind. However, what the results in \ref{fig:psample_spat_mean} demonstrate is that we can use the spatial component included in the model shown in equation \eqref{eqn:model_h1} in order to capture the dependency structure between location and wind speed which cannot be accounted for by the other components in the model.
\par 
At present, the majority of the data is located in the South-West of South Africa. Thus, we are most certain of the mean of the spatial field in this region. In future, the application of the model shown in equation \eqref{eqn:model_h1} to more spatially diverse data would allow for a more effective projection of the spatial field over the the entire domain. However, the ability of the model to capture spatial variation in wind speeds validates the inclusion of a spatial component into our model.
\par
This research was conducted with reference to broader socio-economic objectives which align with the sustainable development goals (SDGs) set out by the United Nations. In particular, our findings contribute toward the achievement of SDG-7, affordable and clean energy, and SDG-13, climate action. By enabling the more widespread adoption of wind energy for power generation, we can reduce South Africa's reliance on environmentally harmful, coal driven power stations. By supplanting fossil fuel based electricity generation with renewable energy, we can ensure that renewable electricity is equitably accessible to all citizens of South Africa.





\subsection*{Credit author statement}
Matthew De Bie: performed writing original draft preparation, data conglomeration and result visualisations. Janet Van Niekerk: performed model fitting application and project conceptualisation. Andriette Bekker performed supervision and validation.

\subsection*{Declaration of competing interests}
The authors declare that they have no known competing financial interests or personal relationships that could have appeared to influence the work reported in this paper.
\subsection*{Acknowledgements}
This paper was made possible by the funding provided by the South African Department of Science and Innovation (DSI), the National Research Foundation (NRF Ref. SRUG2204203865 and Ref: RA171022270376 Grant No: 119109), and the Centre of Excellence in Mathematical and Statistical Sciences (CoE-MaSS) .  The opinions expressed and conclusions arrived at during this paper are those of the authors and are not necessarily to be attributed to the DSI-NRF or CoE-MaSS.

\subsection*{Dataset availability}
The code and raw data used to conduct this analysis is available on GitHub at: \\
\url{}
All data and relevant code related to this article can be found at \url{http://surl.li/hqwrl} , an open-source online data repository hosted at Github \cite{debie2023data}.

\bibliographystyle{apalike}
\bibliography{mastersbibliography_references_3}

\end{document}